\documentclass[aip,jcp,reprint,onecolumn,unsortedaddress,showpacs,12pt]{revtex4-1}

\usepackage{amssymb}
\usepackage{epsfig}
\usepackage{amsmath}
\usepackage{times}
\usepackage{graphicx}%
\usepackage{dcolumn}%
\usepackage{bm}%

\usepackage{color}

\newcommand{\vnabla}{ \vec{\nabla} }

\begin{document}

\title{Chiral selection and frequency response of spiral waves in
reaction-diffusion systems under a chiral electric field}

\author{Bing-Wei Li}
\email{bwli@hznu.edu.cn}
\affiliation{Department of Physics, Hangzhou Normal University, Hangzhou 310036, China}
\affiliation{Department of Physics and Astronomy, Ghent University, Krijgslaan 281, Ghent 9000, Belgium}
\author{Mei-Chun Cai and Hong Zhang}
\affiliation{Zhejiang Institute of Modern Physics and Department of Physics, Zhejiang University, Hangzhou 310027, China}
\author{Alexander V. Panfilov and Hans Dierckx}
\affiliation{Department of Physics and Astronomy, Ghent University, Krijgslaan 281, Ghent 9000, Belgium}


\date{\today}

\begin{abstract}
Chirality is one of the most fundamental properties of many physical, chemical and biological systems. However, the mechanisms underlying the onset and control of chiral symmetry are largely understudied.  We investigate possibility of chirality control in a chemical excitable system (the BZ reaction) by application of a chiral (rotating) electric field using  the Oregonator model.  We find that unlike previous findings, we can achieve the chirality control not only  in the field rotation direction, but also  opposite to it, depending on the field rotation  frequency.  To unravel the mechanism, we further develop a comprehensive theory of frequency synchronization based on the response function approach. We find that this problem can be  described by the Adler equation and show phase-locking phenomena, known as the Arnold tongue.  Our theoretical predictions are in good quantitative agreement with the numerical simulations and provide a solid basis for chirality control in excitable media.
\end{abstract}


\maketitle 


\section{Introduction}
Chirality is a significant property of asymmetry that has been found in several branches of science with notorious examples in fields ranging from particle physics to biological systems
\cite{wagniere}. In the context of pattern formation, a typical self-organized wave pattern bearing chirality is a spiral wave, as it has topological charge (either -1 or +1) that is related to the sense of rotation, e.g., clockwise (CW) or counterclockwise (CCW). Spiral waves have been found in a wide
variety of chemical, physical and biological systems. For instance, they occur in the classical
Belousov-Zhabotinsky (BZ) reaction \cite{win72,ouy96,van01}, on platinum surfaces during the process of catalytic oxidation of carbon monoxide \cite{jak90}, in the liquid crystal \cite{frisch}, during aggregation of \textit{Dictyostelium} discoideum amoebae \cite{lee96}, in the chicken retina \cite{yu12} and in cardiac tissue where they are thought to lead to life-threatening cardiac arrhythmias \cite{dav92}.

To date, much attention has been paid to spiral dynamics as they respond to various external
fields such as \emph{dc} and \emph{ac} electric fields \cite{agladze_elc_jpc,steinbock_elc_prl,krinsky_elc_prl,taboada,schmidt,munuzuri94}, periodic forcing \cite{steibocknat,Braunecpl,mantelpre,linprl,zhangprl}, mechanical deformation \cite{munuzuri_elc_pre,sasha,daniel}, and heterogeneity \cite{zoupre93,hans,biktashevprl10,bwlicol,alonsoprl,defauw}. As reaction-diffusion (RD) systems exhibit mirror symmetry, CCW and CW spiral waves are physically identical and the response of spiral waves with opposite chirality to achiral fields is identical up to mirror symmetry. For example, the sense of drift perpendicular to a constant electric field will change for a spiral wave of opposite chirality \cite{agladze_elc_jpc,steinbock_elc_prl,krinsky_elc_prl}.
The chiral property of spiral waves causes some interesting behaviors, in particular as they respond to a chiral field \cite{nicolis_pnas,miglerprl,knam,ecke_sci}. However this issue to our best knowledge remains to be comparatively less addressed over last decades. Recently, a circularly polarized electric field (CPEF) that possesses chirality was theoretically proposed \cite{chenjcp1} and was implemented in the BZ experiment \cite{jipre}, which allows us to study the response of spiral waves to a chiral electric field in RD systems. Indeed, it has been shown that the CPEF has some pronounce effects on spiral waves \cite{chenjcp2,caimc,bwli13} that were not observed in RD systems subject to achiral fields such as a \emph{dc} or \emph{ac} electric field.

Possibly, one of the most interesting results caused by chiral fields or forces is chiral symmetry breaking, an ubiquitously observed scenario in nature \cite{ecke_sci,miglerprl,nicolis_pnas,carmichael,ricci,ribo,noorduin,toyoda}. Over the past decades, chiral symmetry breaking induced by such chiral fields has received considerable interests from many scientific disciplines \cite{ecke_sci,miglerprl,nicolis_pnas,ribo,noorduin,toyoda}. Different from spontaneous chiral symmetry breaking where the
chiral selection is unpredicted, it was demonstrated that chiral fields not only
cause the breaking of chirality but also could select a desired chirality\cite{ecke_sci,miglerprl,nicolis_pnas,ribo,noorduin,toyoda}, which is closely related  with that of the applied field.
As an example, the chirality of a supramolecular
structure can be selected by the vortex motion and depends on the chirality of the vortex \cite{ribo}. In
Rayleigh-B\'{e}nard convection, chiral symmetry breaking in spiral-defect populations was observed when the system rotated along the vertical axis, and the chirality of the dominant spirals relies on the rotation sense of the system \cite{ecke_sci}. By subjecting a RD system to the CPEF \cite{bwli13}, we recently found that the zero-rotation chiral symmetry between CW and CCW spiral defects breaks and that ordered spiral waves with preferred chirality arise from the spiral turbulence state.  Here too, the preferred chirality was only determined by the chirality of the CPEF \cite{bwli13}.

On the other hand, due to the presence of chiral terms, the frequency response of spiral waves with opposite chirality is different. For example, in the complex Ginzburg-Laudau equation (CGLE) with a broken chiral symmetry breaking term, Nam {\it et al}. \cite{knam} showed that this chiral term would cause a shift in the frequency of spiral waves and the sign of this shift depends on the chirality of the spiral waves. In our recent work \cite{bwli13}, such chirality-dependent frequency response was also observed in RD systems coupled to the CPEF. However, a quantitative description of the chirality-dependent frequency response to the CPEF in such RD systems is still lacking.

In this work, we study the competition of a spiral pair in the Oregonator model for the BZ medium coupled to a CPEF and find that the chirality of the dominant spiral pattern can be changed by  tuning the frequency, without altering the CPEF chirality. This finding differs from the previous results where a close relationship between the chirality of the applied filed and that of the selected entity exists. In order to explain this result, we develop a theory for chirality-dependent frequency response using the response functions approach. The theory predicts phase locking phenomena described by Adler equation.  These  theoretical predictions are in good quantitative agreement with the numerical results.

\section{Model and Methods}

\subsection{Reaction-diffusion model}
Experimentally, electric fields are commonly implemented in the BZ chemical reaction to study their effects on spiral waves \cite{agladze_elc_jpc,steinbock_elc_prl}. Most observations can be well reproduced numerically by a Oregonator model that has been modified to take into account the existence of an electric field. Although the Oregonator model had originally three variables, it can be reduced under some situations (e.g., existence of large time scales between chemical species) to a two-variable model \cite{taboada,schmidt}. Previous studies suggested that in the presence of an electric field, computation results based on the two-component version are coincident with the those on the three-variable version \cite{taboada,schmidt}, but with significant savings in computational time. Due to these considerations, in this work we use the following two-component dimensionless Oregonator model for the BZ medium \cite{schmidt,zykovepl}:
\begin{eqnarray}
 \frac{\partial u}{\partial t}&=& \varepsilon^{-1}\left[u-u^{2}-(fv+\varphi)\frac{u-q}{u+q}\right]+M_{u}\vec{E}\cdot \vnabla u +D_{u}\Delta u ,\\ \frac{\partial v}{\partial t}&=&u-v+M_{v}\vec{E}\cdot \vnabla v+D_{v}\Delta v,
\end{eqnarray}
where fast variable $u$ and slow variable $v$ respectively represent the concentrations of the autocatalytic species HBrO$_{2}$ and the catalyst of the reaction. The small dimensionless parameter $\varepsilon$ represents the ratio of time scales of the dynamics of the fast and slow
variables; $f$ is the stoichiometric parameter and the parameter $q$ is a ratio of chemical reaction rates. The parameter $\varphi$ controls the local excitability of the system. $D_{u}$ and $D_{v}$ stand for  the diffusion coefficients of HBrO$_{2}$ and the catalyst. In what follows, we only consider the diffusion of $u$, as we assume that the reaction takes place in a gel which immobilizes the catalyst, i.e., $D_{v}=0$. The effects of an external electric field are considered through the terms $M_{u}\vec{E}\cdot\vnabla u$ and $M_{v}\vec{E}\cdot\vnabla v$ where $M_{u}$ and $M_{v}$ denotes the mobility of the ions under an electric field. We furthermore assume $M_{u}$ and $M_{v}$ to be proportional to $D_{u}$ and $D_v$, implying $M_{v}=0$. Thus, the applied electric field only affects the fast variable $u$ in our case.

The driving force for chiral selection is implemented as the CPEF $\vec{E}=E\cos(\omega_{f} t)\textbf{i}+E\cos(\omega_{f}
t+\Delta\gamma)\textbf{j}$, where $\textbf{i}, \textbf{j}$ are orthogonal basis vectors in the plane. Such CPEF can be generated experimentally by applying two \textit{ac} electric fields perpendicular to each other and
tuning the phase difference $\Delta\gamma$; $\Delta\gamma=3\pi/2$
($\Delta\gamma=\pi/2$) corresponds to a CCW (CW) CPEF. For more details on the experimental setup, please refer to Ref. \cite{jipre}.

\subsection{Numerical methods}

We numerically integrated Eqs.(1-2) using the explicit Euler method with a spatial step $\Delta x=\Delta y=0.20$ s.u. and a time step $\Delta t=0.002$ t.u..
Through our work, we fix $q=0.002$, $\varepsilon=0.1$, $f=2.0$, and $\varphi=0.01$, as in Ref. \cite{zykovepl}, such that the system is in an excitable regime that supports a rigidly rotating spiral wave. The spiral tip is defined by the intersection point of the isolines of $u=0.20$ and $v=0.05$; the rotation frequency of a spiral wave is calculated via $\omega_{s}=2\pi/T$ where $T$ is the arithmetic mean of the time intervals between two successive maximal values of $u$ in a given point of the medium. In the absence of an electric field, the period of the spiral wave was measured to be $T_{0} = 7.382$, corresponding to a natural rotation frequency $\omega_{0} =  0.851$.

\section{Numerical results}

\subsection{Chiral selection of a spiral pair}

\begin{figure}
\includegraphics[bb=90pt 438pt 507pt 773pt,clip,scale=0.90]{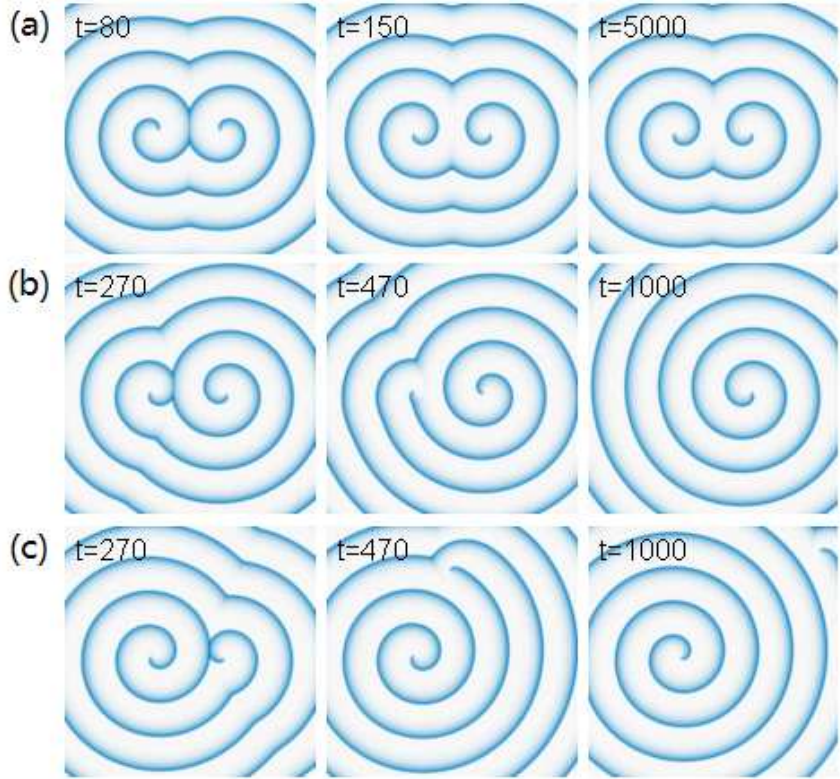} 
\caption{(color online). Chiral symmetry breaking and selection of a spiral pair under a CCW CPEF. (a) $E=0$,  a spiral pair stably rotates. (b) $E=0.05$ and $\omega_{f}=0.87>\omega_{0}$, chiral symmetry breaking occurs, and chirality of the dominant spiral is CCW, the same
as that of the CPEF. (c) similar to (b), but with $\omega_{f}=0.83<\omega_{0}$, the dominant spiral is CW, opposite
to that of the CPEF. Numbers shown in each plot indicate the time. We show the
$v$ variable in each snapshot. The system is composed of $1024\times1024$ grid points.
}
\end{figure}

In this section we investigate the behavior of a spiral pair under a CPEF. To this end, we first initiate a pair of two counter-rotating spiral waves as shown in Fig. 1(a); such pair is invariant under mirror symmetry. We
note that dynamics of a spiral pair has been studied by many authors
\cite{aranson,villarreal_pre,villarreal_prl,schebesch, aranson_prl96}.
For instance, in the framework of CGLE, a spiral pair undergoes a symmetry breaking instability.
Such kind of breaking has been further reported experimentally in the BZ
system \cite{villarreal_pre,brandtstadter} and numerically in the three-component RD systems \cite{schebesch, aranson_prl96}.
In our case such kind of instability does not occur without the electric terms in Eqs. (1-2). Figure 1(b) shows a process of the chiral symmetry instability of the spiral pair after switching on the CCW CPEF with $E=0.05$ and $\omega_{f}=0.87$, slightly larger than the rotation frequency $\omega_{0}$ of a free spiral.
In less than 36 rotations, chiral symmetry is clearly lost [refer to t=270 t.u. in Fig. 1(b)], and then the CCW spiral with the same chirality as the CPEF starts to dominate (refer to t=470 t.u.). Later, the CW spiral is pushed to the boundary and only the CCW spiral wave survives in the system (t=1000 t.u.). Note that in this case the external field selects a spiral with the same chirality as its own, which is consistent with the chiral selection controlled by chiral fields in other systems \cite{ecke_sci,miglerprl,nicolis_pnas,ribo,noorduin,toyoda}.
\begin{figure}
\includegraphics[bb=86pt 630pt 503pt 769pt,clip,scale=1.0]{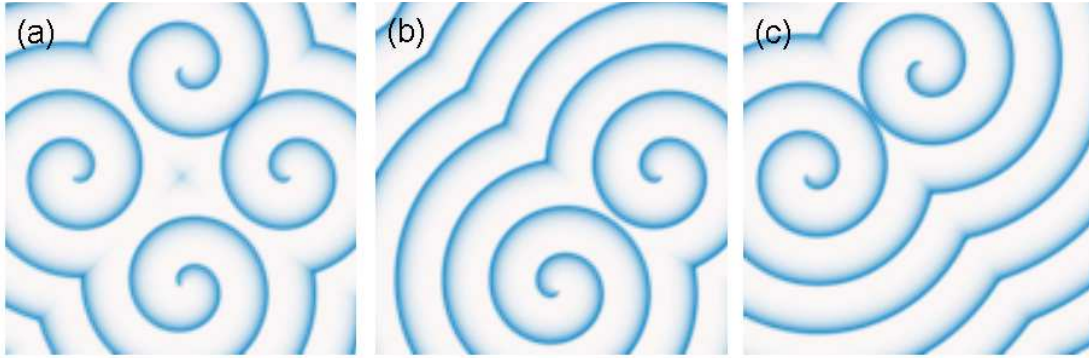} 
\caption{(color online). Chiral selection in a multiple-spiral state caused by a CCW CPEF. (a) $E=0.0$, stable multiple-spiral state. (b) $E=0.05$, $\omega_{f}=0.87$,
CCW spiral waves dominate at $t= 5000$ t.u.. (c) $E=0.05$, $\omega_{f}=0.83$,
CW spiral waves dominate CCW at $t=3000$ t.u..The same system size and parameters as in Fig. 1 were used. }
\end{figure}

However, the opposite chirality (CW) spiral wave can also be selected without changing the chirality of the CPEF. This scenario is illustrated in Fig. 1(c) where we keep the same chirality for the CPEF as in Fig. 1(b), but lower the forcing frequency
$\omega_{f}$ to $0.83<\omega_{0}$. In contrast to Fig. 1(b), we observe
the CW spiral develops and the other one is reduced
to a bare core. Thus, the spiral with the opposite chirality to the CPEF is eventually selected. This scenario is quite different from the previous studies on the chiral symmetry breaking caused by chiral fields, where reversal of the chiral field chirality seems necessary if one needs to select the opposite chiral entity. \cite{ecke_sci,miglerprl,nicolis_pnas,ribo,noorduin,toyoda}.
\begin{figure}
\includegraphics[bb=50pt 185pt 530pt 643pt,clip,scale=0.85]{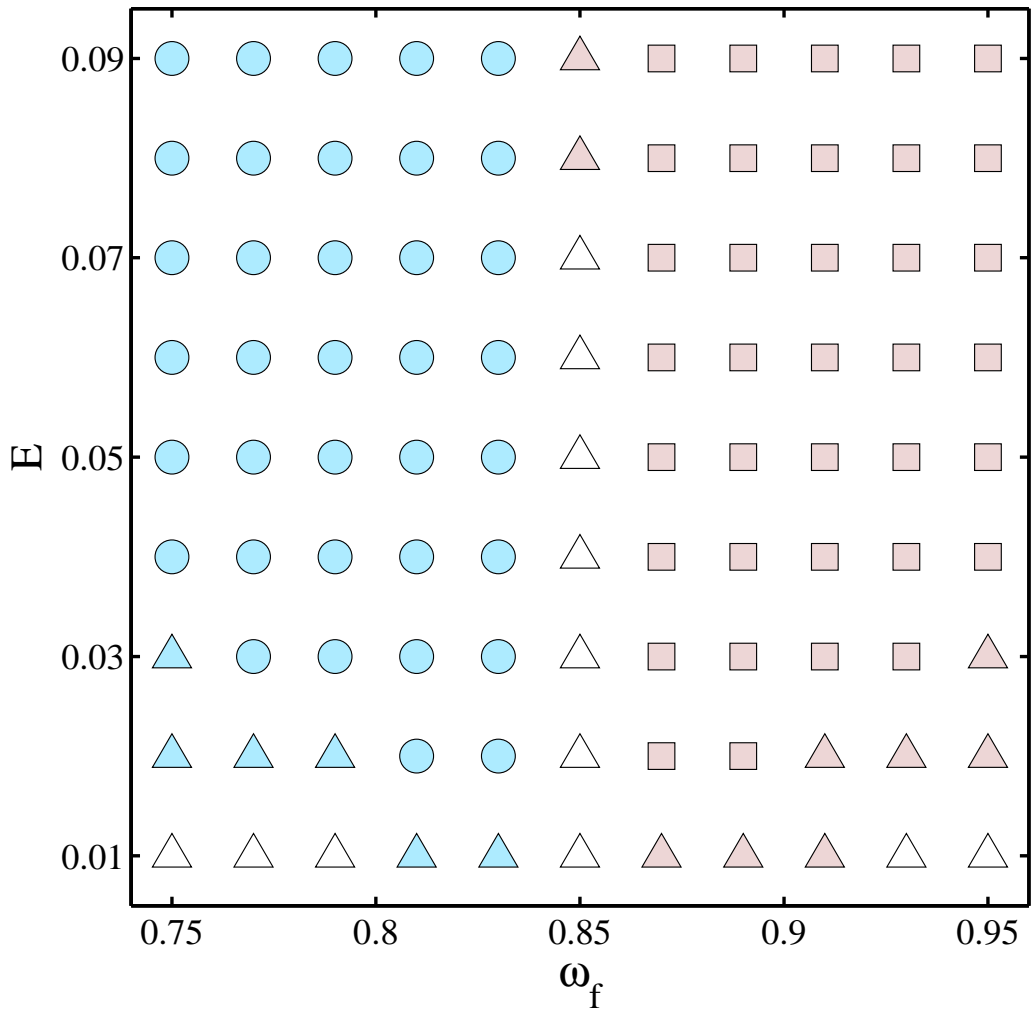} %
\caption{(color online). Phase diagram of chiral selection controlled by the CCW CPEF. Shaded circles and squares denote the CW and CCW dominant spiral waves, respectively; shaded triangles mean that the CCW and CW spiral still coexist but chiral symmetry breaking can be observed. Open triangle denotes the coexistence of the CCW and the CW spiral waves without obvious chiral symmetry breaking at the end of the simulation time $t_{tot}=1000$ t.u.. To compute this diagram, we use the same system size as in Fig. 1.}
\end{figure}

\subsection{Chiral selection of multi-spiral states}
The present findings also differ from the spontaneous symmetry breaking of spiral pairs found previously where chiral selection of the dominant spiral is unpredictable which is sensitive to the many factors such as the initial distance between the spiral core\cite{aranson,villarreal_pre,villarreal_prl,schebesch, aranson_prl96,brandtstadter}. However, the forced chiral symmetry breaking caused by the CPEF presented in this work [both in Fig. 1(b) and (c)] is quite robust and almost insensitive to the initial orientation or inter-spiral distance. For example, the chiral symmetry breaking observed above is not only limited to a spiral pair and it can also occur in a state with multiple spiral waves, as illustrated in Fig. 2. Similar to the interaction of a spiral pair, we find CCW spiral waves would dominate CW spiral
waves when we apply CCW CPEF with $\omega_{f}=0.87 > \omega_0$ and CW spirals finally dominate
CCW ones as we change the forcing frequency to $\omega_{f}=0.83 < \omega_0$.

\begin{figure}
\includegraphics[width = \textwidth]{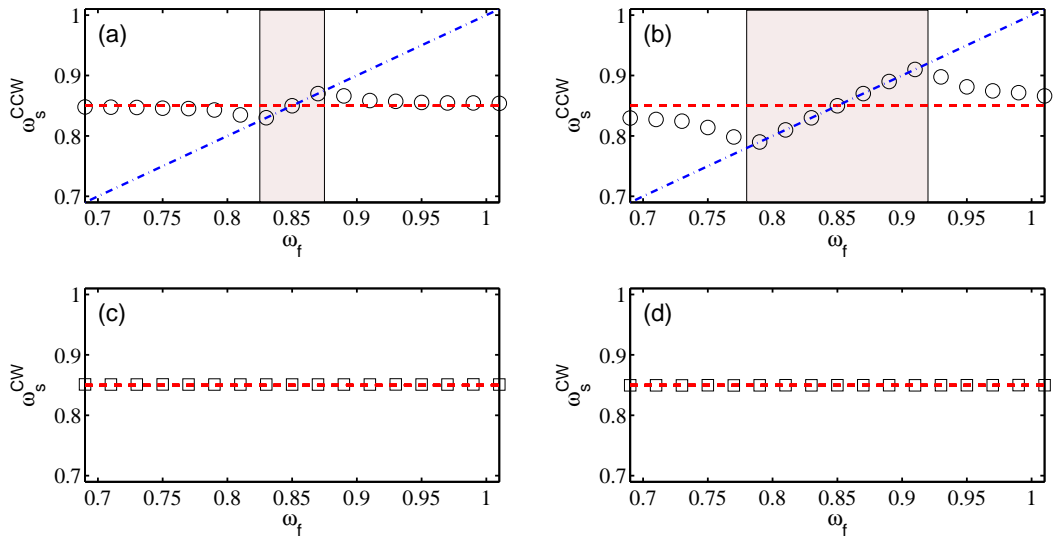}
\caption{(color online). Spiral frequency response to the CCW CPEF. (a-b) The frequency of a CCW spiral wave as a function of the forcing frequency $\omega_{f}$ for $E=0.02$ and $E=0.05$, respectively. (c-d) The frequency of a CW spiral
as a function of $\omega_{f}$ for $E=0.02$ and $E=0.05$, respectively. The red dashed lines represent $\omega_{0}=0.851$. In (a-b), the shaded region denotes the synchronized region and the blue dots lines mean $\omega_{s}=\omega_{f}$. To calculate the frequency (period), we use the system with $256\times256$ grid points.}
\end{figure}

\subsection{Phase diagram for chiral selection}

A systematic study shows that such chiral symmetry breaking and pattern selection can be achieved in a broad parameter range. We summarize the results in the phase diagram of $E$ versus $\omega_{f}$ in Fig. 3. After a spiral pair is created ($t=40$ t.u.), the CPEF is switched on and the evolution is followed until the simulation was ended ($t= t_e=1040$ t.u.). Different final states are coded with different markers in Fig. 3. First, when no obvious chiral breaking is noticed, open triangles are drawn. This happened near resonance ($\omega_f \approx \omega_0$) and at the lower corners of the diagram where $|\omega_f - \omega_0| / E$ is large. Secondly, we drew shaded circles (squares) when a single dominant CW (CCW) spiral pattern survived. As in Fig. 1(a-b), these regions correspond to where $\omega_f$ is slightly smaller (bigger) than $\omega_0$. Third, between the fully selective and non-selective regions of the phase diagram, a border zone exists (colored triangles), where chiral symmetry was broken at $t=t_e$, without achieving a selection of single-chirality spiral waves.

\subsection{Frequency response of spiral waves to a CPEF}

The time-averaged spiral frequency $\omega_{s}$ is measured (refer to the section of the numerical method) and its dependence on the forcing frequency $\omega_{f}$ for two intensities $E$ is plotted in Fig. 4 (a-b) for the CCW spiral and Fig. 4(c-d) for the CW one. From these plots, we find that there is a chirality-dependent frequency response of spiral waves to the CPEF. Specifically, the CCW spiral with the same chirality as the CPEF, is able to keep the pace with the CPEF, which is particularly obvious if the frequency mismatch between
$\omega_{f}$ and $\omega_{0}$, denoted by $\Delta \omega=\omega_{f}-\omega_0$, is small. Once this happens, $\omega^{CCW}_{s}$ would be altered to keep the same value as $\omega_{f}$ and frequency synchronization, also known as phase-locking, is observed. The synchronization region (color shaded) is extended as we increase the strength of the CPEF [compared Fig. 4(a) to 4(b)].  However, rotation frequency of the spiral waves with opposite chirality to the CPEF, seems to be hardly affected by the frequency and the intensity of the CPEF. The rotation frequency of CW spirals stays close to $\omega_{0}$, as seen from Figs. 4(c-d).

\section{Mechanism for chiral selection and frequency synchronization}

\subsection{Selection of spiral waves by frequency shift}
The chirality-dependent frequency response to the CPEF is the underlying cause for the chiral selection observed in Figs. (1-3) as we explain below from the viewpoint of the wave competition. When we apply a CCW CPEF with the frequency that is quite close to but a little larger than $\omega_{0}$, i.e., $\omega_{f}>\omega_{0}$, from Fig. 4 we know the rotation frequency of the CCW spiral with the same chirality as the CPEF would be increased due to synchronization: i.e., $\omega^{CCW}_{s}=\omega_{f}$; while for the CW spiral, its frequency $\omega^{CW}_{s}$ seems not changed and thus $\omega^{CW} \approx \omega_{0}$.
This causes a frequency shift between CCW and CW spiral waves when a spiral pair is affected by the CPEF. Due to the competition rule that in excitable media the faster one always win the slower one, we find at the end the CCW spiral dominates the CW one in Fig. 1(b). E.g. the faster source may push its wave tail closer to the other spiral's core, which is eventually directly exposed to the spiral wave of the higher frequency. As a result the slower rotating wave will drift, and may be annihilated at the medium boundary. The scenario is also very similar if we apply the CPEF with $\omega_{f}<\omega_{0}$, e.g., see Fig. 1(c). This competition also explains the scenarios witnessed in Figs. (2-3).

When we apply $\omega_{f}$ that is almost equal to $\omega_{0}$, we will get
$\omega^{CCW}_{s} \approx \omega^{CW}_{s}$, and under this situation, a spiral pair will still be  stable as the case without the CPEF. If the frequency shift between spiral waves caused by the CPEF is extremely small, we need a much longer time to observe chiral symmetry breaking, which
explains the narrow zone at $\omega_f \approx \omega_0$ in Fig. 3 where chiral symmetry was intact at the end of simulation time.

\subsection{Theoretical description of frequency synchronization using response function theory}

In order to look into the nature and origin of the chirality-dependent frequency response, especially frequency synchronization, in further detail, we below derive a phase equation based on the based on the singular perturbation theory \cite{biktashev94,biktashevapre03,biktashevapre09,biktashevapre10,dxspiral,dierckx2,keener86,henrypre02,mikhailov94} around an unperturbed spiral wave, employing critical adjoint eigenfunctions which are also known as response functions \cite{biktashev94,biktashevapre03,biktashevapre09,biktashevapre10,dxspiral,dierckx2}. From this phase equation, we analytically find the conditions under which a spiral wave will synchronize with the CPEF, as we now proceed to show.

\begin{figure}
\includegraphics[bb=316pt 44pt 715pt 441pt,clip,scale=0.60]{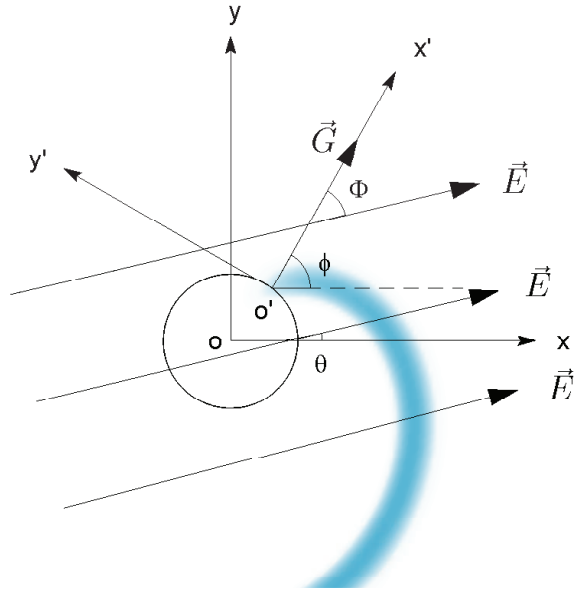}
\caption{(color online). A scheme of defining references and angles for our derivation. $xoy$ is fixed in the lab reference, while $x'o'y'$ co-rotates with the spiral at the instantaneous frequency $\omega$. $o'x'$ is chosen as parallel with the vector $\vec{G}=\vnabla u / ||\vnabla u ||$ at the spiral tip. For then, $\Phi=\phi(t)-\theta(t)$ where $\theta(t)=\theta_{0}+\omega_{f}t$ and $\phi(t)=\phi_{0}+\omega_{0}t+\tilde{\phi}(t)$ denote the rotation phases of the electric field and the spiral wave. The shaded region denotes the shape of the spiral wave.  \label{frames}}
\end{figure}

We start our derivation of the phase-locking equation by rewriting Eqs.(1-2) in a matrix form,
\begin{equation}
\partial_t \textbf{u}=\hat{\textbf{D}}\triangle\textbf{u}+\textbf{F(u)} + \textbf{h}, \label{RDE}
\end{equation}
where in our case $\textbf{u}=(u,v)^{T}$, $\textbf{F}=(\varepsilon^{-1}(u-u^{2}-(fv+\varphi)(u-q)/(u+q)),u-v)^{T}$ and $\textbf{h}=\vec{E}\cdot\hat{\textbf{M}}\vnabla \textbf{u}$ is assumed to be a small perturbation. Furthermore, $\hat{\textbf{D}}$ and  $\hat{\textbf{M}}$ are constant diffusion and mobility matrices, given by
\begin{align}
\hat{\textbf{D}}&=\left(\begin{array}{cc}
                                              D_{u} & 0 \\
                                              0 &  D_{v}
                                            \end{array} \right), &
\hat{\textbf{M}}&=\left(\begin{array}{cc}
                                              M_{u} & 0 \\
                                              0 &  M_{v}
                                            \end{array} \right).
\end{align}

Next, we introduce two frames of reference as shown in Fig. 5. The laboratory frame is denoted by $(x,y,t)$, while the spiral frame that co-rotates with the spiral wave at the instantaneous frequency $\omega$ is denoted by $(x',y',t')$.
To measure the spiral's rotation phase, we introduce the reference vector $\vec{G}=\vnabla u / ||\vnabla u ||$ at the spiral tip. The phase $\phi$ is then the oriented angle between the $X$-direction and $\vec{G}$. We will choose the co-rotating frame such that $\vec{G}$ is aligned with the positive X'-axis at all times.

Furthermore, we denote by $\sigma$ the chirality of the spiral wave: $\sigma=+1$ for CCW and $\sigma=-1$ for CW rotation. (As before, we assume the rotation of the CPEF is always CCW, i.e., $\omega_{f}>0$.)

According to the response function theory used in Refs. \cite{biktashev94,biktashevapre03,biktashevapre09,biktashevapre10,dxspiral,dierckx2}, a small perturbation $\textbf{h}$ acts on a robust spiral pattern by causing a translational and rotational shift. In particular, the phase angle will evolve as $\phi(t) = \phi_{0}+\sigma\omega_0 t + \tilde{\phi}(t)$ with phase correction $\tilde{\phi}(t)$. Hence, the instantaneous rotation frequency changes to $\dot{\phi}(t) =\sigma\omega_{0} + \tilde{\omega}(t)$ where $\tilde{\omega}(t) = \dot{\tilde{\phi}}(t)$ is a convolution of $\textbf{h}$ with the so-called rotational response function \cite{dierckx2}, i.e.,
\begin{eqnarray}
\tilde{\omega}&=&
\int_{\mathbb{R}^{2}}{\textbf{W}^{(0)}(x',y')^{H}\textbf{h}(x',y',t')dx'dy'.}\label{omg}
\end{eqnarray}
Here $(.)^H$ denotes Hermitian conjugation of a column vector of state-variables and $\textbf{W}^{(0)}$ is the rotational response function. Mathematically, it is the adjoint zero mode to the linearized operator associated to Eq. \eqref{RDE} \cite{keener86, biktashevapre09, dierckx2}. We use the normalization as in Ref. \cite{biktashevapre09} such that
\begin{equation}
\int_{\mathbb{R}^{2}} \textbf{W}^{(0)}(X,Y)^{H}\partial_{\Theta}\textbf{u}_0(X,Y)dXdY= -1,
\end{equation}
to avoid the appearance of a minus sign in Eq. \eqref{omg}. Here, $\textbf{u}_0(X,Y)$ is the time-independent unperturbed spiral wave solution to Eq. \eqref{RDE} in the co-rotating frame with polar coordinate $\Theta$. For a given RD model with differentiable reaction kinetics, $\textbf{W}^{(0)}$ can be numerically computed, see e.g. Refs.\cite{henrypre02, biktashevapre09}.
In the present case, the perturbation is the CPEF, i.e., $\textbf{h}=\vec{E}\cdot\hat{\textbf{M}}\vnabla \textbf{u}$, which can be expressed in the spiral frame ($x'o'y'$) as
\begin{eqnarray}
\textbf{h}&=&E^{x'}\hat{\textbf{M}}\partial_{x'}\textbf{u}+E^{y'}\hat{\textbf{M}}\partial_{y'}\textbf{u} 
= E^{x'}\hat{\textbf{M}}\partial_{x'}\textbf{u}_{0}+E^{y'}\hat{\textbf{P}}\partial_{y'}\textbf{u}_{0}+O(E^{2}). \label{pert}
\end{eqnarray}
Here $E^{x'}$ ($E^{y'}$) is the component of the electric field $\vec{E}$ along $x'$ ($y'$) and a first approximation is made, i.e., $\textbf{u}(x',y',t')=\textbf{u}_{0}(x',y') + \tilde{\textbf{u}}(x',y',t'),$ where $\tilde{\textbf{u}}$ is of the same order as $E$. Substituting  Eq. (\ref{pert}) to Eq. (\ref{omg}), we have,
\begin{eqnarray}
\tilde{\omega}&=& \dot{\tilde\phi}=E^{x'}M_{x'}^{0}+E^{y'}M_{y'}^{0}+O(E^{2}), \label{peq1} \end{eqnarray}
where $M_{x'}^{0}$ and $M_{y'}^{0}$ are time-independent constants  given by the overlap integrals

\begin{eqnarray}
M_{x'}^{0}=\int_{\mathbb{R}^{2}}{\textbf{W}^{(0)}(x',y')^{H}\hat{\textbf{M}}\partial_{x'}\textbf{u}_{0}(x',y')dx'dy'}, M_{y'}^{0}=\int_{\mathbb{R}^{2}}{\textbf{W}^{(0)}(x',y')^{H}\hat{\textbf{M}}\partial_{y'}\textbf{u}_{0}(x',y')dx'dy'}.
\end{eqnarray}

Let us now denote the experimentally accessible angle of $\vec{G}$ relative to the electric field as $\Phi(t) = \phi(t) - \theta(t)$, which yields  $E^{x'} = E \cos \Phi$ and $E^{y'} = - E \sin \Phi$. If we further define $M_{x'}^{0}=A\cos\alpha$ and $M_{y'}^{0}=A\sin\alpha$ such that
\begin{align}
A&=\sqrt{(M_{x'}^{0})^{2}+(M_{y'}^{0})^{2}},& \tan\alpha&=M_{y'}^{0}/M_{x'}^{0},
\end{align}
Eq. \eqref{peq1} can be written as
  \begin{eqnarray}
\dot{\tilde{\omega}}=\dot{\tilde\phi} = EA \cos(\Phi + \alpha)+O(E^{2}). \label{peq2}
\end{eqnarray}
Recalling that $\Phi(t) = \phi(t) - \theta(t)$, we note that $\dot{\Phi} = (\sigma\omega_0-\omega_f) + \dot{\tilde{\phi}}$, which finally produces the phase equation, up to linear order in the field intensity $E$:
  \begin{eqnarray}
\dot{\Phi}(t) =-\Delta\omega+ EA \cos(\Phi + \alpha),\label{peq}
\end{eqnarray}
where $\Delta\omega=\omega_{f}-\sigma\omega_{0}$.

Interestingly, we note that Eq. (\ref{peq}) has the same form as with phase synchronization of oscillators driven by a small periodic force \cite{kuramoto,pikovsky} where it is often called the Adler equation \cite{adler}. This equation is served to study phase-locking phenomena in diverse natural or engineered systems\cite{adler, kuramoto}.
Furthermore, the coefficients $A$ and $\alpha$ can be found from numerical computation, as we will proceed to show.

\begin{figure}
\includegraphics[bb=280pt 190pt 737pt 638pt,clip,scale=0.80]{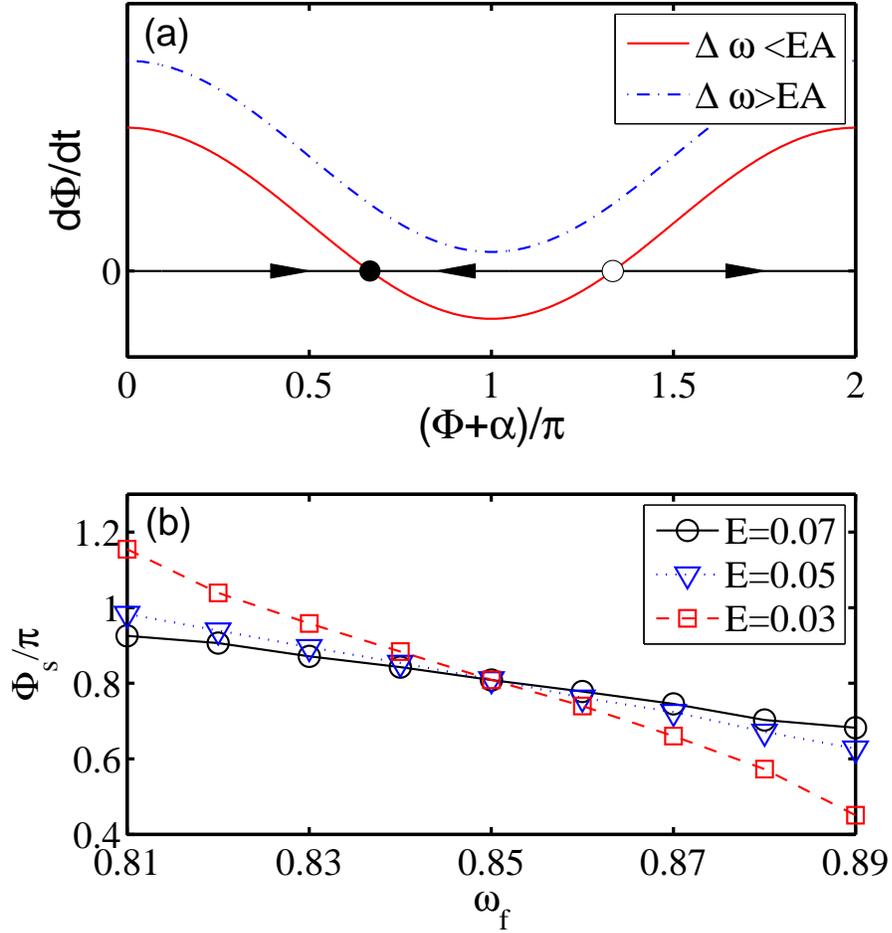} 
\caption{(color online). (a) Plot of $\dot\Phi$ versus $(\Phi+\alpha)$. The solid black dot represents the stable solution (fixed point) while the open circle denotes the unstable one. Arrows denote the flow direction. (b) Dependence of numerically measured $\Phi_{s}$ on $\omega_{f}$ for three different intensities of the electric field. }
\end{figure}

Let us first, however, discuss how frequency synchronization follows
from Eq. \eqref{peq}. Since $|\cos(x)| \leq 1$ for all real-valued arguments $x$,
in the case of the same rotation between spiral waves and the CPEF
the one-dimensional dynamical system Eq. \eqref{peq} possesses two equilibrium points
$\Phi_\pm = - \alpha \pm \arccos\left(\frac{\Delta\omega}{EA}\right)$
whenever $|\omega_0-\omega_f |  < EA$. From Fig. 6(a), it is observed that only the equilibrium point which has $0<\Phi +\alpha<\pi$ is stable; we will denote it as
\begin{equation}
\Phi_{s}=-\alpha+\arccos\left(\frac{\Delta\omega}{EA}\right). \label{phis}
\end{equation}
Hence, a unique phase-locked spiral state exists as long as
\begin{equation}
E > E_* = \frac{\left| \omega_0-\omega_f \right| }{A}.  \label{synbdy}
\end{equation}
Such phase-locking region is known as an Arnold tongue; for the current system and order of calculation in $E$, it has a triangular shape.

If the spiral wave and the CPEF rotate in an opposite way, the phase equation can be written explicitly as $\dot{\Phi}=-(\omega_{0}+\omega_{f})+EA\cos(\Phi+\alpha)$. Since we work in the regime of $EA<< (\omega_{0}+\omega_{f})$, we expect no synchronization in this case, which is consistent with the observation in Fig. 4(c-d).

\begin{figure}
\includegraphics[bb=280pt 190pt 737pt 638pt,clip,scale=0.80]{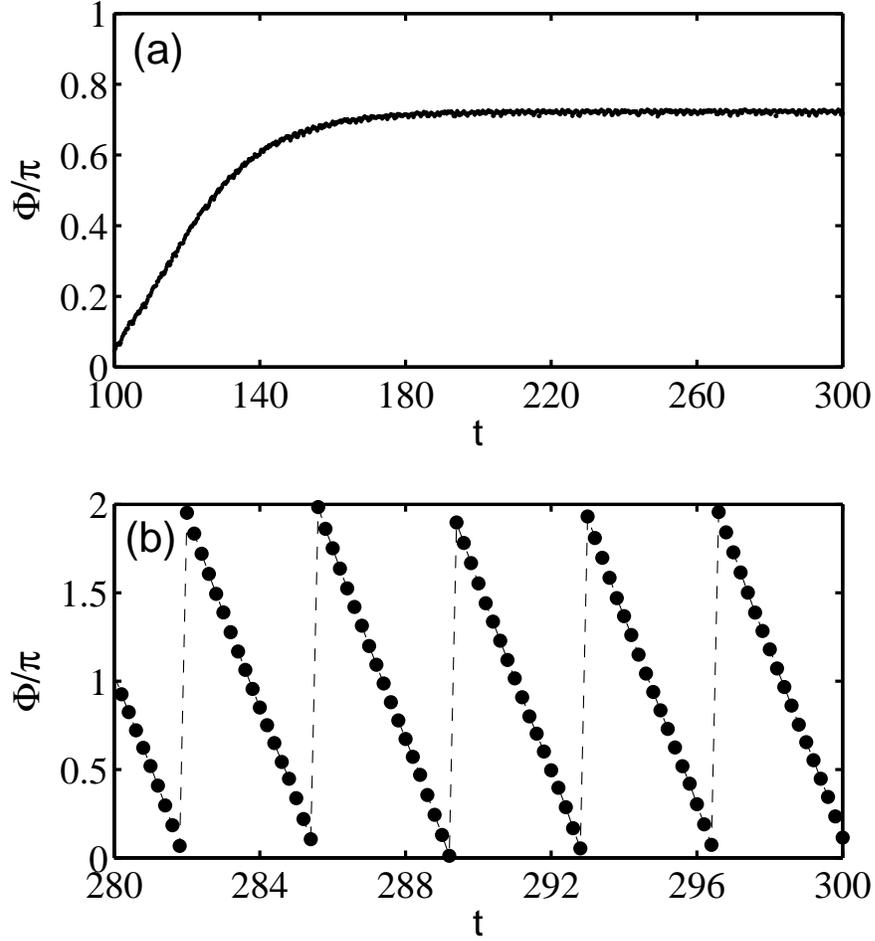} 
\caption{(a) $\Phi$ changes as time for $E=0.05$ and $\omega_{f}=0.87$ showing the phase locking for large $t$. Both the spiral and the CPEF are CCW. (b) Same as (a) but for the opposite chirality, e.g., CW spiral and CCW CPEF. }
\end{figure}

\subsection{Quantitative results using response functions}

\begin{figure}
\includegraphics[bb=280pt 190pt 737pt 638pt,clip,scale=0.80]{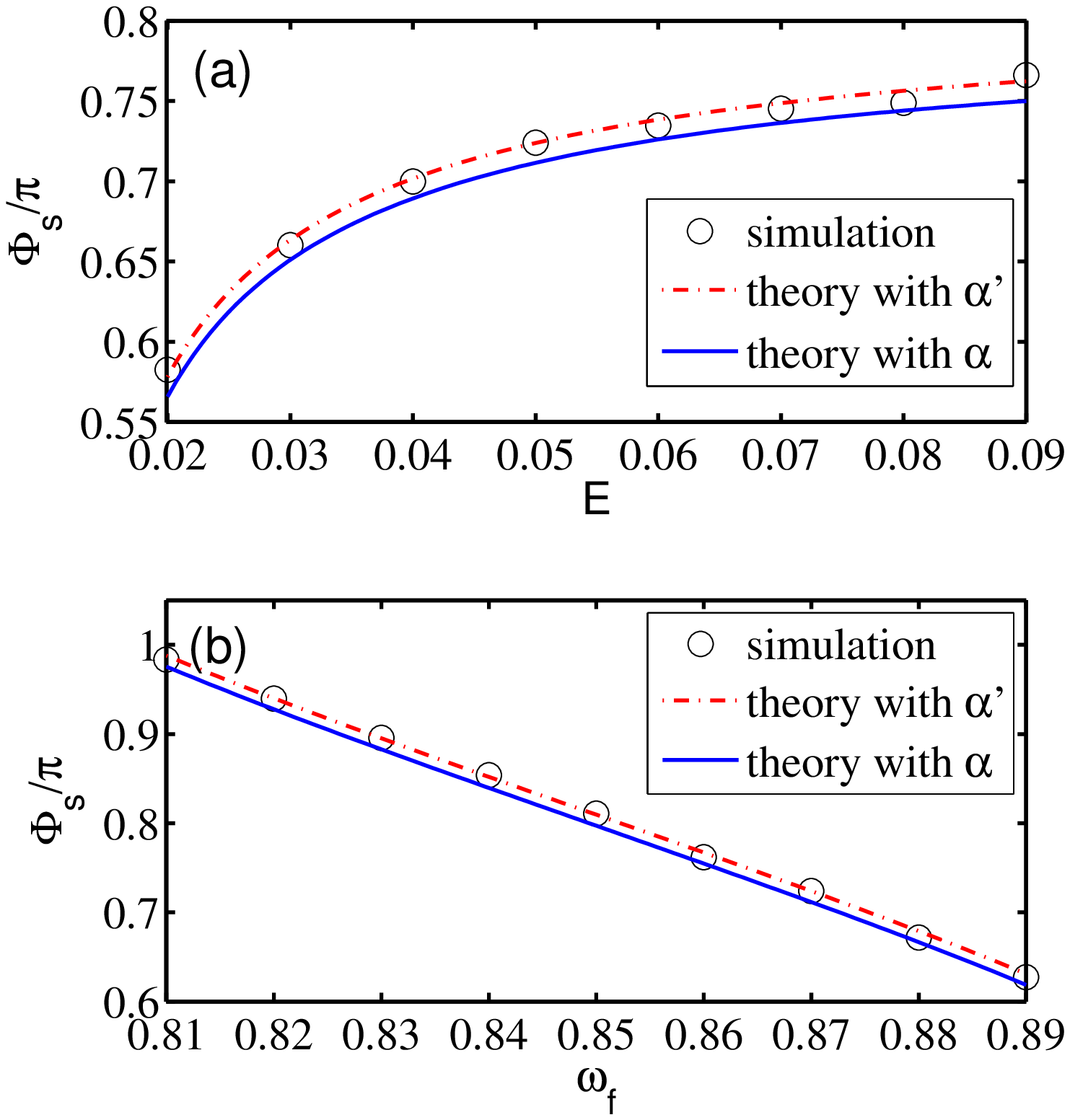} 
\caption{(color online). A comparison of theory with numerical simulations. (a) $\Phi_{s}$ as a function of $E$ for $\omega_{f}=0.87$; (b) $\Phi_{s}$ as a function of $\omega_{f}$ for $E=0.05$. Theoretical predictions from Eq. (\ref{phis}) with $\alpha'=-0.9722$ [calculated from Eq. (\ref{phis}) and Fig. 6(b)] and $\alpha=-0.9334$ (calculated from the response functions) are used. The same system size as in Fig. 4 is used to compute these plots. }
\end{figure}

Finally, we will quantitatively compare the theory and numerical experiments for the value of the phase-locked angle and the width of the Arnold tongue. These quantities only depend on the parameters $A$ and $\alpha$, which are fully determined by the parameters of the RD system without the electric field. They can be directly calculated with response functions by using the freely available software \textsc{dxspiral} \cite{biktashevapre09,biktashevapre10,dxspiral}, to which we added reaction kinetics for the Oregonator model. With the parameters used in our paper and a polar grid of radius $R= 10 $ with $N_r = 240, N_\theta= 256$, we find $A = 1.505$. In the frame where $x'$ is aligned with $\vec{G}$, we find $M^\theta_{x'} = 0.8955$, $M^\theta_{y'} = -1.2094$, such that $\alpha = -0.9334 = -53.5^\circ$.

To compare our theory and simulations, we first note that at resonance ($\Delta \omega=0$), the relation Eq. \eqref{phis} predicts that the phase-locked angle found between $\vec{E}$, $\vec{G}$ should equal $\Phi_s(\Delta \omega = 0) = - \alpha + \pi/2.$
In our numerical experiment, $\Phi_s(\Delta \omega = 0)$ was measured to be $145.7^\circ$, as seen in Fig. 6(b) where we plot $\Phi_{s}$ as function of $\omega_{f}$, yielding $\alpha' = -0.9722 =-55.70^\circ$. (We here use $\alpha'$ to distinguish from the value of $\alpha$ that is directly calculated from the response functions.)  This value $\alpha'$ is closely matched by our response function calculation above (i.e., $\alpha$), with an error about $2.2^\circ$.

Figure 7 shows the typical evolution for $\Phi$, the spiral phase relative to the electric field,  in the case of phase-locking [Fig. 7(a)] or no phase-locking [Fig. 7(b)]. Panel 7(a) shows the change of $\Phi$ as time elapses for the CCW spiral in the presence of the CCW CPEF with $E=0.05$ and $\omega_{f}=0.87$. One observes that $\Phi$ eventually reaches a constant, indicating frequency synchronization (i.e., phase-locking). In Fig. 7(b) with the same parameters but with CW spiral waves, however, $\Phi$ changes periodically with period close to $2\pi/(\omega_{f}+\omega_{0})$. The system behavior in both panels is consistent with Eq. (\ref{peq}).

In the synchronization case, $\Phi_{s}$ is determined by Eq. (\ref{phis}), implying that both $\Delta\omega$ and $E$ affect the phase-locked angle $\Phi_{s}$.
In Fig. 8(a), we predict the dependence of $\Phi_{s}$ on $E$ given $\omega_{f}=0.87$ (i.e., $\Delta\omega=0.02$)
using Eq. (\ref{phis}) with $A$ and above mentioned values of $\alpha$ (and $\alpha'$).
One observes that the simulation results agree well with the theoretical prediction from Eq. (\ref{phis}). Similarly, the graph in Fig. 8(b) of $\Phi_{s}$ as a function of $\omega_{f}$ given $E=0.05$ also shows a nice correspondence between simulations and theory. Note that a linear dependency of $\Phi_{s}$ on $\omega_{f}$ is found since $\arccos(x) \approx \pi/2-x$ in the frequency synchronization regime shown here.
\begin{figure}
\includegraphics[bb=59pt 270pt 535pt 557pt,clip,scale=0.7]{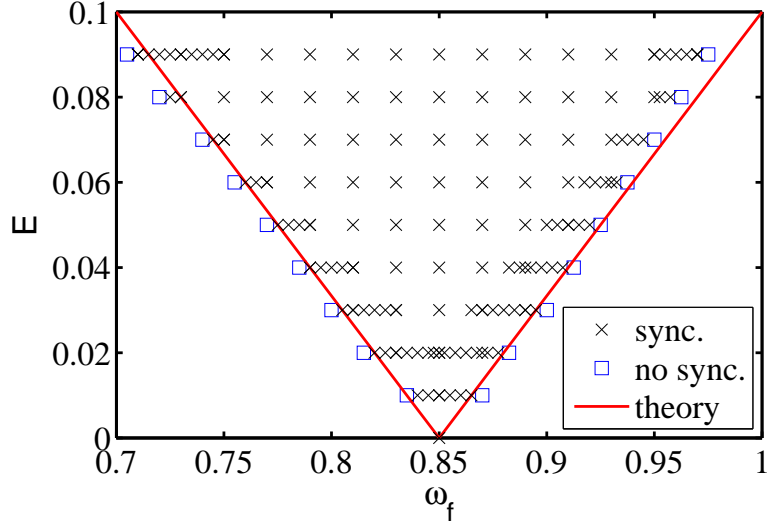} 
\caption{(color online). Arnold synchronized tongue. The crosses  and square denote the numerical simulations where synchronization and unsynchronization behaviors are observed; the red line shows the boundary derived from Eq. (\ref{peq}) with $A=1.505$ (without free parameters). To compute this plot, we use the same system size as in Fig. 4.  }
\end{figure}

To give a complete comparison between these predictions and numerical results, we present in Fig. 9 the Arnold tongue for synchronization in the parameter space of $\omega_{f}$ and $E$. In this figure, crosses denote the numerical results of the synchronization. For each value of $E$ shown, we started simulations in the synchronization zone and increased or decreased $\omega_f$ until synchronization could no longer be established. The couples ($E, \omega_{f}$) where synchronization failed are represented by the blue squares in Fig. 9. For comparison, the red line denotes the synchronization boundary that is derived from Eq. (\ref{synbdy}). We find that a good agreement is achieved for the small $E$. Note that, for $E>0.05$ deviations are visible which may be captured by extending our linear theory to high orders in $E$ in the future.

Finally, it is noted that synchronous mechanism plays an important role in the chiral selection shown in  Figs. (1-2), however, such kind of chiral symmetry breaking is not only limited to the synchronous region as seen by comparing the Arnold tongue (Fig. 9) with the phase diagram (Fig. 3). This is because even in the not fully synchronized region, the CPEF can also cause a chirality-dependent frequency response as illustrated in Fig. 4 (outside the shaded region). For example, in the unsynchronized regime but still close to synchronous regime, the spiral frequency would be still larger (smaller) than $\omega_{0}$ if we apply the CPEF with $\omega_{f}$ larger (smaller) than $\omega_{0}$. Therefore even in the non-synchronous region, we could also observe chiral symmetry breaking and pattern selection.
\section{Discussion}

By studying the instability of spiral pairs in the Oregonator model subject to the CPEF, we demonstrated and quantitatively analyzed the chiral selection of spiral pairs controlled by a chiral electric field. Our results are different from the previous findings. On one hand, prior to our work, most of works showed that the chirality of the dominant pattern seemed fully determined by that of the applied field, however, our work showed another possibility in chemical media that the dominant spiral pattern could be the same as or opposite to the chirality of the CPEF. On the other hand, in previous work, chirality-dependent frequency response, especially frequency synchronization between spiral waves and the applied CPEF was only discussed in the phenomenological level and the dynamical mechanism was unclear. In the present work, based on the response function theory, we found the coupling of spiral waves and the applied CPEF can be transformed to a phase equation that governs the evolution of the angle of spiral orientation relative to the electrical field. From this equation, we could make several quantitative predications that were validated by our numerical simulations. It is worth noting that our findings were quite robust and insensitive to the specific models. For example, we also checked the main results in Barkley's model \cite{barkley91} and found similar results. From this point of view, our present work allows us to understand better about the interaction between spiral waves and the CPEF and provides a solid basis for chirality control in excitable media.

We would like to point out that the proposed theory for frequency synchronization is applicable for the rigidly rotating spiral waves only. However,  previous work showed that  such CPEF induced  frequency synchronization or phase-locking  also occurs  for the meandering spiral waves \cite{chenjcp2}. It would be important  to extend our theory in future  to describe such phase locking phenomenon for the meandering case.

Finally, the findings present in this work can be directly tested in chemical experiments, since the CPEF has been recently realized in the BZ system \cite{jipre}. Compared to the spiral turbulence state \cite{jipre}, we believe that it is easier to observe chiral symmetry breaking and pattern selection by considering a stable spiral pair under the CPEF. For, to observe chiral symmetry breaking in the spiral turbulence state, in addition to consider the wave competition between spiral waves, we also need to consider the issue of stabilization of the unstable spirals.  This would take longer time and require stronger intensity of the CPEF. The latter factor is quite important in the implementation of the CPEF in the laboratory because the stronger intensity of the CPEF could cause increased heat production which leads to some serious problems (e.g., higher or uncontrollable excitability) for the experimental set-up\cite{jipre}.

\section{Conclusion}

In summary, we have investigated chiral symmetry breaking and pattern selection in RD systems (Oregonator model) coupled to the CPEF. We have shown that the chirality of dominant pattern can be well controlled by the applied electric field in the synchronous (or unsynchronous) regime. More interestingly, we showed that we can select an opposite chiral pattern by tuning the forcing frequency instead of changing the chirality of the CPEF. We attribute this scenario to the chirality-dependent frequency response to the CPEF, which can be quantitatively described by the Adler equation that we have originally derived using response function theory. Our predictions agree well with the numerical results. Finally, considering the recent realization of the CPEF in the BZ system and that our results are robust throughout numerical simulations, we believe that our findings presented here are highly likely to be observed in the laboratory.

\section*{Acknowledgments}
The authors are grateful to the team that developed \textsc{dxspiral} and made it publicly available. B.W.L. thanks the financial support from the National Nature
Science Foundation of China under Grant No. 11205039 and the Visiting Scholars Program of Hangzhou Normal University. H.D. was founded by FWO-Flanders. This work was also partially supported by the funds from Hangzhou City for supporting Hangzhou-City Quantum Information and Quantum Optics Innovation Research Team.


%
%

%



\end{document}